\documentclass[twocolumn,aps,pra,groupedaddress,showpacs]{revtex4}
\usepackage{epsfig,amssymb}
\def\comment#1{}\def\labell#1{\label{#1}}
\begin{document}
\title{Information rate of a waveguide} \author{Vittorio
  Giovannetti$^1$, Seth Lloyd$^{1,2}$, Lorenzo Maccone$^1$, and
  Jeffrey H. Shapiro$^1$}\affiliation{$^1$Massachusetts Institute of
  Technology -- Research Laboratory of Electronics\\$^2$Massachusetts
  Institute of Technology -- Department of Mechanical Engineering\\ 77
  Massachusetts Ave., Cambridge, MA 02139, USA}

\begin{abstract}
  We calculate the communication capacity of a broadband
  electromagnetic waveguide as a function of its spatial dimensions
  and input power.  We analyze the two cases in which either all the
  available modes or only a single directional mode are employed.  The
  results are compared with those for the free space bosonic channel.
\end{abstract}
 \maketitle 

In the analysis of electromagnetic communication channels using
quantum information, emphasis has been placed on free space
communication protocols, in which waves propagate unconstrained: a
good {\em summa} of all the obtained results can be found in
{\cite{caves,yuen}}. Here we will focus on constrained communication
lines, such as optical fibers or radio waveguides, in the lossless
limit.  Although the spatial mode structure for free space propagation
between a pair of apertures has long been understood {\cite{slep}},
its near-field modes only approximate the lossless behavior of ideal
waveguides. In contrast, since the spatial properties of the waveguide
modes are always well-defined, we will be able to derive the exact
dependence of the information rate on the system parameters, e.g., the
power $P$ and the waveguide cross-sectional area $A$, obtaining
results that closely resemble the ones described in
{\cite{caves,moore}} for the free space channel.  Moreover, as will be
discussed in detail, our derivation resolves some of the open issues
connected with the optimization of the multimode communication
protocols.

We start by describing the waveguide communication channel in
Sec.~\ref{s:ch} and calculate the rate in Sec.~\ref{s:proc}. In
particular, Secs.~\ref{s:math} and \ref{s:narrow} are devoted to the
regimes of multiple modes and single directional mode respectively.
The discussion and the comparison with prior results is given in
Sect.~\ref{s:disc}.

\begin{figure}[hb]
\begin{center}
\epsfxsize=.8\hsize\leavevmode\epsffile{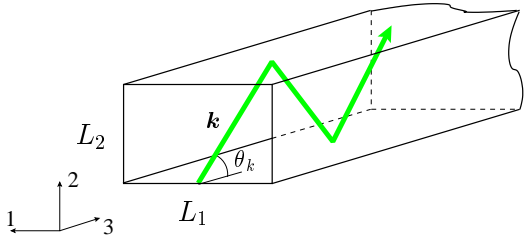}
\end{center}
\caption{Description of the ideal metallic waveguide. The modes (TE or
  TM) with wave vector {\boldmath$k$} propagate in the positive
  longitudinal direction bouncing off the perfectly reflecting walls
  of the waveguide.}  \labell{f:ch}\end{figure}
 
\section{The channel}\labell{s:ch}
Although guided-wave optical communications are normally carried out
using dielectric waveguides, metallic waveguides provide a simpler
mode structure for deriving the broadband information rate. In the
ideal, lossless case that we consider, such waveguides confine the
electromagnetic field into a finite region of space by means of
perfectly-reflecting boundaries.  In this paper we will analyze in
detail the rectangular cross-section case with transverse spatial
dimensions $L_1$ and $L_2$ described in Fig.~\ref{f:ch}, even though
the procedure can be readily extended to other configurations. In a
hollow waveguide the transverse electromagnetic modes (TEM)
customarily used in free space communications do not propagate.  They
are replaced by the transverse electric (TE) and the transverse
magnetic (TM) modes in which only the electric or magnetic terms
respectively possess null longitudinal components. These modes are
characterized by the wave vectors {\cite{haus}} \vbox{
    \begin{eqnarray} \mbox{\boldmath $k$}=\left(\frac{\pi\:n_1}{L_1},
\frac{\pi\:n_2}{L_2},k_3\right)
&& \left\{\begin{array}{ll}n_1,n_2=1, 2,\cdots&\mbox{TE}\\\\
n_1,n_2=0, 1, 2,\cdots&\mbox{TM}\\
(n_1=n_2=0 \mbox{ excluded})&\end{array}\right.\nonumber\\
&&\labell{modi}
\end{eqnarray}}
where the discretization of the transverse components derives from the
boundary conditions at the waveguide walls and where the longitudinal
component $k_3$ is a positive quantity because we are considering only
modes propagating from the sender to the receiver.  By introducing the
transmission time $\cal T$ (i.e. the time interval in which the sender
operates the channel), the longitudinal parameter $k_3$ can be
discretized using periodic boundary conditions. In particular, a mode
with wave vector {\boldmath $k$} bounces off the waveguide walls (see
Fig.~\ref{f:ch}) so that it propagates across the transmission line
with a longitudinal speed (group velocity) $c\cos\theta\equiv c
k_3/|\mbox{\boldmath $k$}|$. This means that all the photons of that
mode used in the transmission can be ideally enclosed in a box of
longitudinal length $c {\cal T} \cos\theta$: assuming periodic
boundary conditions $k_3$ can be discretized as   $2\pi n_3/(c{\cal
T}\cos\theta)$ with $n_3$ a positive integer. This relation introduces
a nonlinearity in the dependence of the mode frequencies
$\omega_k(n_1,n_2,n_3)\equiv c|${\boldmath $k$}$|$ on the parameters
$n_1$, $n_2$ and $n_3$, i.e.
\begin{eqnarray}
\frac{\omega_{\mbox{\boldmath $k$}}(n_1,n_2,n_3)}c= f\left(\frac{\pi n_1}{L_1},\frac{\pi
    n_2}{L_2},\frac{2\pi n_3}{c{\cal T}}\right),\labell{a}
\end{eqnarray}
where
\begin{eqnarray}
f({\mbox{\boldmath 
    $x$}})\equiv\frac{2\left(x_1^2+ x_2^2\right)}
{-x_3+ \sqrt{x_3^2+4\left(x_1^2+x_2^2\right)}}\;.
\;\labell{defdif}
\end{eqnarray}
As will be discussed in Sect.~\ref{s:math}, this is the main
difference between our approach and the free space calculation
performed in {\cite{caves}}.

\section{The communication rate}\labell{s:proc}
The communication rate $R$ is the maximum number of bits per second
that can be transmitted through the channel, and is given by the
capacity (i.e.  the maximum of the mutual information between the
input and the output of the channel) divided by the transmission time
$\cal T$. The capacity can be estimated from a quantum mechanical
analysis of the communication in which each symbol $\theta$,
transmitted with probability density $p(\theta)$, is associated with a
quantum state $\rho(\theta)$ of the Hilbert space $\cal H$ of the
media used in the communication process (in our case the
electromagnetic field). Without any constraints, the infinite
dimensions of the input space $\cal H$ of the system under
consideration can accommodate an arbitrary amount of information and
the capacity would diverge.  Physically it is, thus, sensible to
introduce an energy constraint on the accessible input
states~\cite{yuen}. In particular, we consider the following limit on
the available average power (i.e., energy transmitted per unit time)
\begin{eqnarray} P=\frac{\mbox{Tr}[H\rho]}{\cal T} \;\labell{power},
\end{eqnarray} 
where $\rho=\int d\theta\:p(\theta)\rho(\theta)$ is the average
message sent through the channel (i.e. the electromagnetic field state
at the input), and $H$ is the Hamiltonian of the modes, i.e.
\begin{eqnarray} H=\sum_{\mbox{\scriptsize\boldmath $k$}
    ,\epsilon}H_{{\mbox{\scriptsize\boldmath $k$},\epsilon}}\;,\quad
  H_{{\mbox{\scriptsize\boldmath $k$},\epsilon}}\equiv
  \hbar\omega_{k}\;a_{\mbox{\scriptsize\boldmath $k$},\epsilon}^\dag
  a_{\mbox{\scriptsize\boldmath $k$},\epsilon}\;\labell{hamilt},
\end{eqnarray}
with $\epsilon=$TE,TM and $a_{\mbox{\scriptsize\boldmath$k$},\epsilon}$
being the annihilation operator of the mode $\epsilon$ with wave
vector {\boldmath $k$} and commutator
         \begin{eqnarray}
\left[a_{\mbox{\scriptsize\boldmath
$k$},\epsilon},a^\dag_{\mbox{\scriptsize\boldmath
$k$}',\epsilon'}\right]=\delta_{\epsilon\epsilon'}\delta_{\mbox{\scriptsize\boldmath
$kk$}'}\;\labell{commut}.
\end{eqnarray}
In order to calculate the capacity of the channel under consideration,
i.e. the maximization of the mutual information under the
constraint~(\ref{power}), we need the infinite dimensional
extension~\cite{yuen} of the Holevo theorem~\cite{hol1} which, in the
noiseless case, gives an upper bound to the capacity in terms of the
maximal input von Neumann entropy. As discussed in Ref.~\cite{yuen},
this upper bound is achievable, so that the rate is
\begin{eqnarray} R=\max_{\rho}\frac{S(\rho)}{\cal
    T} \;\labell{rate},
\end{eqnarray}
where $S(\rho)=-$Tr$[\rho\log_2\rho]$ is the Von Neumann entropy and
the maximum is taken over all the possible density matrices $\rho$ of
the modes employed in the transmission, which satisfy
Eq.~(\ref{power}). Notice that Eq.~(\ref{rate}) might be seen as an
instance of the Holevo-Schumacher-Westmoreland theorem~\cite{hsw}, but
for the case under consideration was first derived by
Yuen-Ozawa~\cite{yuen}. The maximization of Eq.~(\ref{rate}) under the
constraint (\ref{power}) can be performed by means of a variational
principle: the maximum is reached for the $\rho$ that satisfies
      \begin{eqnarray} \delta\left\{\frac{S(\rho)}{\cal
T}-\frac{\lambda}{\ln 2}\frac{\mbox{Tr}[H\rho]}{\cal
T}-\frac{\lambda'}{\ln 2}\mbox{Tr}[\rho]\right\}=0 \;\labell{variat},
\end{eqnarray}
where $\lambda$ and $\lambda'$ are the two Lagrange multipliers that
derive from the power constraint (\ref{power}) and from the
normalization condition on $\rho$, respectively (the factor $\ln 2$
has been inserted so that all calculations can be performed using
natural logarithms). Using standard techniques (see for instance
{\cite{bekenstein}}), it is possible to show that Eq.~(\ref{variat})
is satisfied by the density matrix
 \begin{eqnarray}
\rho_{max}=\frac{e^{-\lambda_0 H}}{Z(\lambda_0)}
\;\labell{rhom},
\end{eqnarray}
where $Z(\lambda)\equiv$Tr$[e^{-\lambda H}]$ is the partition function
and $\lambda_0$ is determined by the equation
\begin{eqnarray}
P=-\left.\frac{\partial}{\partial\lambda}\left(\frac{\ln Z(\lambda)}{\cal
    T}\right) \right|_{\lambda_0}
\;\labell{lambda}.
\end{eqnarray}
Using this solution, the maximum rate in bits per unit time is finally
given by
 \begin{eqnarray}
R=\frac 1{\ln 2}\left(\lambda_0 P+\frac{\ln Z(\lambda_0)}{\cal T}\right)
\;\labell{sol}.
\end{eqnarray}
To obtain an explicit expression for $R$, we thus only need to
evaluate the partition function $Z(\lambda)$ for the Hamiltonian
(\ref{hamilt}). In the two following sections we will undertake such
endeavor for two different communication scenarios.

\subsection{Multimode communication}\labell{s:math}
In this section we calculate the rate $R$ when all the wave vectors
that propagate in the positive longitudinal direction (from the sender
to the receiver) are employed in the communication.

Since modes with different {\boldmath $k$} or different $\epsilon$ are
independent, the partition function factorizes in product of single
mode partition functions $Z${\boldmath$_k$}$_{,\epsilon}(\lambda)$ so
that
       \begin{eqnarray}
\ln Z(\lambda)=\sum_{\mbox{\scriptsize\boldmath
$k$},\epsilon}\ln Z_{\mbox{\scriptsize\boldmath $k$},
\epsilon}(\lambda) \;\labell{z1},
\end{eqnarray}
with
\begin{eqnarray}
Z_{\mbox{\scriptsize\boldmath
    $k$},\epsilon}(\lambda)\equiv\mbox{Tr}[e^{-\lambda H_{\mbox{\scriptsize\boldmath $k$},\epsilon}}]= 
\frac 1{1-e^{-\lambda\hbar\omega_k}} 
\;\labell{z2}.
\end{eqnarray}
Substituting (\ref{z2}) into (\ref{z1}) one can compute $Z(\lambda)$
by summing over the allowed values of $n_1$, $n_2$ and $n_3$. Since we
are interested in the stationary information rate, we should take the
limit ${\cal T}\to\infty$ which allows the summation over $n_3$ to be
replaced with an integral. Even with this simplification, the
calculation is quite demanding and is postponed to the final
paragraphs of this subsection for the sake of readability.  Here we
consider the simpler high power/high cross-section regime, defined by
the condition
 \begin{eqnarray} \gamma\equiv\frac
  {AP}{c^2\hbar}\gg 1 \;\labell{condiz},
\end{eqnarray}
where $A=L_1L_2$ is the cross sectional area of the waveguide. In this
regime too the summations over $n_1$ and $n_2$ reduce to integrals
and, apart from corrections of order $1/\gamma$, Eq.~(\ref{z1})
becomes
\begin{eqnarray}
\ln Z(\lambda)\simeq\frac{gAc{\cal T}}{2\pi^3}\int_V d{\mbox{\boldmath
    $x$}}\;\ln\left[\frac 1{1-e^{-\lambda\hbar c
      f({\mbox{\scriptsize\boldmath 
    $x$}})}}\right]
\;\labell{z5},
\end{eqnarray}
where the volume integral must be performed on the subspace $V$ of
positive $x_j$ and $f(${\boldmath $x$}$)$ is defined in
Eq.~(\ref{defdif}). The parameter $g=2$ in Eq.~(\ref{z5}) counts the
different species of modes, TE and TM in this case. It plays the same
role as the polarization degeneracy in the free space propagation of
electromagnetic waves. By performing a change of integration variables
and using the integral of Eq.~(\ref{integr}) of App.~\ref{s:int},
Eq.~(\ref{z5}) reduces to
\begin{eqnarray}
\ln Z(\lambda)=\frac {g\pi^2}{240}\frac{A{\cal T}}{\hbar^3\lambda^3c^2}
\;\labell{f}.
\end{eqnarray}
Substituting this result in Eq.~(\ref{lambda}) gives
\begin{eqnarray}
\lambda_0=\left(\frac{g\pi^2}{80}\frac{A}{P\hbar^3c^2}\right)^{\frac 14}
\;\labell{la},
\end{eqnarray}
which, through Eq.~(\ref{sol}) implies the following maximum rate
\begin{eqnarray} R=\frac 4{3\ln 2}\left(\frac{g\pi^2}{80}\frac
    A{c^2}\right)^{\frac 14}\left(\frac P\hbar\right)^{\frac 34}
  \;\labell{risult1}.
\end{eqnarray}

\begin{figure}[hb]
\begin{center}
\epsfxsize=1.\hsize\leavevmode\epsffile{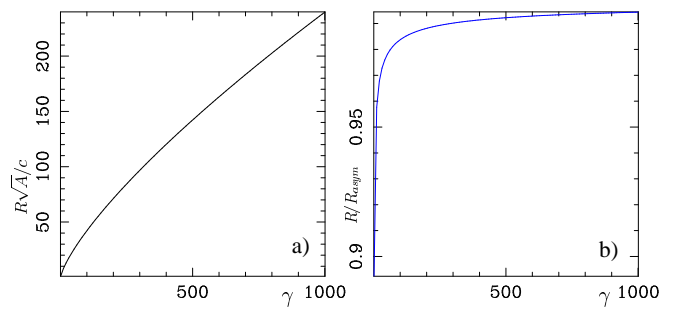}
\end{center}
\caption{Numerical plots. a) Plot of the rate  $R$ given by
  Eq.~(\ref{sol1}) as a function of the dimensionless parameter
  $\gamma$. b) Comparison between the same solution and the asymptotic
  behavior $R_{asym}$ of Eq.~(\ref{soluz1}): the ratio between these
  two quantities tends to one for $\gamma\gg 1$. }
\labell{f:num}\end{figure}

\paragraph*{Numerical results.---}\labell{s:cal}
When condition (\ref{condiz}) does not apply, the summations over
$n_1$ and $n_2$ in the definition of $\ln Z(\lambda)$ cannot be
performed analytically. In this case we can resort to numerical
evaluation of the rate. For the sake of simplicity we will consider a
waveguide with square cross section, i.e.  $L_1=L_2$. Remembering that
not all the values of $n_1$ and $n_2$ contribute both to the TE and to
the TM modes, the summation of Eq.~(\ref{z1}) can be written as
\begin{eqnarray}
\ln Z(\lambda)=\frac{c{\cal T}}{\sqrt{A}}\;{\cal
  W}\!\left(\frac{\pi\lambda\hbar c}{\sqrt{A}}\right)\labell{zeta5}, 
\end{eqnarray}
with
\begin{eqnarray}
{\cal W}(\beta)\equiv \sum_{n_1,n_2=1}^\infty
  F_{n_1,n_2}(\beta)+
  \sum_{n_1=1}^\infty F_{n_1,0}(\beta)\;\labell{z3},
\end{eqnarray}
where
\begin{eqnarray}
F_{n_1,n_2}(\beta)\labell{z4}\equiv\int_0^\infty
  dx\;\ln\left[\frac 1{1-e^{-\beta f(n_1,n_2,x)}}\right]\;.
\end{eqnarray}
Using Eq.~(\ref{sol}), one can show that $R\;\sqrt{A}/c$ is a function
only of the dimensionless parameter $\gamma$, defined in
(\ref{condiz}). In fact Eq.~(\ref{sol}) becomes
\begin{eqnarray}
R=\frac{c}{\ln 2\sqrt{A}}\left[\frac{\gamma\beta_0}\pi+{\cal W}(\beta_0)\right]
\;\labell{sol1},
\end{eqnarray}
where $\beta_0$ is the dimensionless quantity determined by the
condition (\ref{lambda}), i.e., the solution of 
\begin{eqnarray}
\left.\frac{\partial}{\partial\beta}{\cal W}(\beta)\right|_{\beta_0}=-\frac\gamma{\pi}
\;\labell{z7}.
\end{eqnarray}
In Fig.~\ref{f:num} the numerical evaluation of the rate $R$ is
reported. Notice that in the limit $\gamma\gg 1$, the solution
(\ref{sol1}) approaches the asymptotic behavior
\begin{eqnarray}
\frac{R\;\sqrt{A}}c\to\frac 4{3\ln 2}\left(\frac{g\pi^2}{80}\right)^{\frac
  14}\gamma^{\frac 34}
\;\labell{soluz1}
\end{eqnarray} 
which corresponds to the high-power/high cross-sectional limit
solution of Eq.~(\ref{risult1}) discussed previously.

The dimensionless parameter $\gamma$ that identifies the onset of the
asymptotic regime for $R\sqrt{A}/c$ has a relatively simple physical
interpretation. From the density matrix, Eq.~(\ref{rhom}), we see that
the occupancy probabilities are highest for the lowest frequency
(lowest photon energy) modes. From the exact formulation for the
partition function, Eq.~(\ref{z1}), we see that the triple-integral
approximation in Eq.~(\ref{z5}) will be valid when the occupancy
probabilities change very little between modes with adjacent energy
levels. The largest such photon-energy spacing occurs for the
lowest-frequency modes, and, roughly speaking, is equal to $\hbar
c/\sqrt{A}$, i.e., the photon-energy of the waveguide's cutoff
frequency. If we concentrate all of the sender's average power into
the lowest frequency mode, the resulting power spectral density will
be approximately $P\sqrt{A}/c$, and hence $\gamma$ equals this
spectrum measured in units of the photon energy $\hbar c/\sqrt{A}$ of
the lowest-order mode. The condition $\gamma\gg 1$ then guarantees the
desired smooth behavior of the occupancy probabilities, because
Eq.~(\ref{rhom}) implies that high-photon-number occupancy of the
lowest-order mode will, of necessity, be accompanied by similar
occupancy of other low frequency modes. 

This analysis clarifies the regime of applicability of the
approximation (\ref{risult1}) and underlines the importance of the
quantity $\gamma$ in the definition of maximum rate.

\subsection{Single directional mode communication}\labell{s:narrow}
In this section we calculate the rate $R$ when the wideband
transmission is limited to using a single direction of the wave vector
{\boldmath $k$}. We will specify this direction assigning the polar
angle $\theta=\arccos(k_3/|${\boldmath$k$}$|)$ and the azimuth angle
$\varphi=\arctan(k_2/k_1)$. In terms of the discretization parameters
$n_1$, $n_2$ and $n_3$, these conditions become
\begin{eqnarray}
\frac{n_2}{n_1}&=&\frac{L_2}{L_1}\tan\varphi\;\labell{cc2}\\
\left(\frac{\pi n_1}{L_1}\right)^2+\left(\frac{\pi
    n_2}{L_2}\right)^2&=&
\frac{\sin^2\theta}{\cos^4\theta}\left(\frac{2\pi n_3}{c{\cal T}}\right)^2
\;\labell{cc1},
\end{eqnarray}
where the nonlinear relation of Eq.~(\ref{a}) was used in deriving
Eq.~(\ref{cc1}). In this case, only those modes with {\boldmath$k$}
compatible with the chosen direction contribute to the partition
function sum (\ref{z1}), i.e.
\begin{widetext}
\begin{eqnarray}
&&\ln Z(\lambda)=\sum_{n_3=0}^\infty\sum_{n_1,n_2=1}^\infty 
\ln\left[\frac 1{1-e^{-2\pi\lambda n_3/({\cal
        T}\cos^2\theta)}}\right]
\ 
\delta_{n_2,n_1\tan\varphi{L_2}/{L_1}}\quad
\delta_{n_1,2n_3{L_1\sin\theta}/[{c{\cal T}(1+\tan^2\varphi)^{1/2}\cos^2\theta}]}
\;\labell{z},
\end{eqnarray}
where, for the sake of simplicity only the TM mode has been considered
and where the two Kronecker deltas take into account the conditions
(\ref{cc2}) and (\ref{cc1}). Again working in the high power/high
cross-sectional regime (\ref{condiz}), the summations can be replaced
with integrals and the Kronecker deltas become Dirac delta functions,
so that we find
\begin{eqnarray}
\!\!\ln Z(\lambda)\simeq\frac{c{\cal T}}{2\pi}\int_V \!d{\mbox{\boldmath
    $x$}}\;\ln\left[\frac 1{1-e^{-\lambda\hbar c
      x_3/\cos^2\theta}}\right]
\delta\left(x_2-x_1\tan\varphi\right)\;\delta\left(x_1-x_3
\frac{\sin\theta}{\sqrt{1+\tan^2\varphi}\cos^2\theta}\right)
=\frac{\pi{\cal
    T}\cos^2\theta}{12\lambda\hbar}
\;\labell{l1},
\end{eqnarray}
\end{widetext}
where $V$ is again the subspace of positive components of $x_j$.
Substituting this result into Eqs.~(\ref{lambda}) and (\ref{sol}), we
now find that the maximum rate is
\begin{eqnarray}
R=\frac{\cos\theta}{\ln 2}\sqrt{\frac{\pi P}{3\hbar}}
\;\labell{risultc}.
\end{eqnarray}
If both the TE and TM modes of the chosen direction were used for the
transmission, then a factor $\sqrt{2}$ would appear in
Eq.~(\ref{risultc}).

\section{Discussion}\labell{s:disc}
In the previous section we calculated the maximum rate $R$ for
information transmission through an ideal metallic waveguide under an
average input power constraint. In particular we found that when the
sender is using all the available modes, $R$ scales as
$A^{1/4}P^{3/4}$, as reported in Eq.~(\ref{risult1}). This scaling is
reminiscent of the free space communication one {\cite{caves}}. The
main difference between the two cases is that, for waveguides all the
positively propagating {\boldmath$k$} vectors actually reach the
receiver thanks to the reflecting walls of the waveguide. For
frequency-$\omega$ propagation over an $L$-m-long free space path
between identical circular apertures of diameter $D$, there are
approximately $[D^2\omega/(8cL)]^2$ low-loss propagation modes, per
polarization state {\cite{slep}}. In essence, the low-loss modes
represent propagation angles that lie within the solid angle subtended
by the receiver at the sender. These low-loss modes can be accounted
for by introducing a factor of $\sqrt{\sin{\theta_{max}}}$
($\theta_{max}$ being the channel angular aperture as seen by the
sender) in the free space rate --- see Sect.~VI A in {\cite{caves}}.
The lossy free space modes that do not satisfy the preceding angular
subtense condition can be used for communication, but their analysis
requires inclusion of an accompanying noise source, which is mandated
by the quantum theory of loss. As noted at the end of this section,
finding the capacity of the lossy propagation channel is a
considerably more difficult problem.

Apart from these physical considerations, a technical difference
between our calculation and the free space analysis is also evident.
In deriving their result, the Authors of {\cite{caves}} instead of
using Eq.~(\ref{rate}), calculate the maximum rate as
\begin{eqnarray}
R=\max_{\rho_{{\mbox{\scriptsize\boldmath
        $k$},\epsilon}}}\sum_{{{\mbox{\scriptsize\boldmath
        $k$}},\epsilon}} \frac{S(\rho_{{\mbox{\scriptsize\boldmath
        $k$},\epsilon}})}{\cal T}\cos\theta
\;\labell{cas},
\end{eqnarray}
where $\theta$ is the polar angle of the mode wave vector
{\boldmath$k$} and the maximum is performed over the mode states
$\rho${\boldmath$_k$}$_{,\epsilon}$ ($\epsilon$ here counts the
different polarizations of free space electromagnetic waves).  The
presence of the term $\cos\theta$ is introduced in the sum to take
into account the difference in the longitudinal speed of mode
propagation. [In our calculation it is the nonlinear Eq.~(\ref{a})
that takes care of this]. Accordingly, the power constraint is
calculated as
\begin{eqnarray}
P=\sum_{{\mbox{\scriptsize\boldmath
        $k$},\epsilon}}\frac{\mbox{Tr}[H_{{\mbox{\scriptsize\boldmath
        $k$},\epsilon}}\rho_{{\mbox{\scriptsize\boldmath
        $k$},\epsilon}}]}{\cal T}\cos\theta
\;\labell{power2}.
\end{eqnarray}
This procedure assumes implicitly that the maximum communication rate
is achieved by a global state of the input modes which is unentangled
over {\boldmath $k$}. This assumption is correct as can be seen (at
least for the waveguide communication protocol studied here) from the
factorized form of the state in Eq.~(\ref{rhom}). In order to compare
the two approaches, we have calculated the maximum rate of the
waveguide using Eqs.~(\ref{cas}) and (\ref{power2}) in place of
(\ref{rate}) and (\ref{power}). The results are, predictably, similar
to the ones reported in Sec.~\ref{s:math}, even though the numerical
factor differs: in fact, for multimode communication and high power
regime, we now find
\begin{eqnarray}
R=\frac 4{3\ln 2}\left(\frac{g\pi^2}{120}\frac
    A{c^2}\right)^{\frac 14}\left(\frac P\hbar\right)^{\frac 34}
\;\labell{risult4},
\end{eqnarray}
smaller than Eq.~(\ref{risult1}) by a $(3/2)^{1/4}$ factor, which
derives from the particular choice of maximization of
Eq.~(\ref{cas}). 

If we consider the case of a single directional mode, on the other
hand, $R$ scales as $\cos\theta\;P^{1/2}$ (see Eq.~(\ref{risultc})):
this is the same result obtained in the free space propagation
{\cite{caves,yuen}}, apart from the $\cos\theta$ factor that takes
into account the decrease in longitudinal propagation speed of the
field due to the reflections at the waveguide walls.

All the results discussed in this paper have been obtained in the
lossless case, in which all the photons injected into the waveguide
arrive to the receiver. In the presence of loss, the calculation
procedure complicates noticeably: the capacity is no more simply given
by the entropy of the initial state, but by the Holevo quantity, which
is not known to be additive over successive uses of the channel
{\cite{hsw,chuang,c}}.  Finally, we have not considered the presence of
prior entanglement shared between the sender and the receiver. In this
case the rate $R$ can be doubled by using the super-dense coding
protocol {\cite{sdc}}.  The calculation of the entanglement assisted
capacity for the single directional mode case in the presence of loss
was given in {\cite{nostro}}, following the procedure of
{\cite{loro}}.

\section{Conclusions}\labell{s:concl}
In conclusion we have calculated the maximum communication rate for a
perfect waveguide in the two regimes of multimode and single
directional-mode communication. A comparison with the known results on
free space communication schemes has been given.

This work was funded by the ARDA, NRO, NSF, and by ARO under a MURI
program.

\appendix
\section{}\labell{s:int}
In this appendix the integration needed for Eq.~(\ref{z5}) is given.

After performing the change of variables $y_j=2\lambda\hbar c x_j$
($j=1,2,3$), the integral in Eq.~(\ref{z5}) becomes
\begin{eqnarray}
&&\int_0^\infty dy_1\int_0^\infty dy_2\int_0^\infty dy_3\;
\ln\left[\frac 1{1-e^{-f({\mbox{\scriptsize\boldmath 
    $y$}})}}\right]\labell{a1}\\\nonumber&&
=\frac\pi 2\int_0^\infty dx\;x\int_0^\infty
dy\;\ln\left[\frac
  1{1-e^{2x^2/\left(y-\sqrt{y^2+4x^2}\right)}}\right]
\;,
\end{eqnarray}
where in the right hand term polar coordinates have been employed in
the $(y_1,y_2)$ plane.  Changing to polar coordinates also in the
plane spanned by $(2x,y)$, the integral becomes
 \begin{eqnarray}
\!\! \int_0^{\frac\pi 2}d\phi\;\frac{\pi\cos\phi}{(1+\sin\phi)^3}
\int_0^\infty dr\;r^2\ln\left[\frac 1{1-e^{-r}}\right]=\frac{\pi^5}{120}
\labell{integr}.
\end{eqnarray}

 \end{document}